# Computational Chemotaxis in Ants and Bacteria over Dynamic Environments

Vitorino Ramos, Carlos Fernandes, Agostinho C. Rosa and Ajith Abraham

*Abstract*— **Chemotaxis can be defined as an innate behavioural response by an organism to a directional stimulus, in which bacteria, and other single-cell or multicellular organisms direct their movements according to certain chemicals in their environment. This is important for bacteria to find food (e.g., glucose) by swimming towards the highest concentration of food molecules, or to flee from poisons. Based on self-organized computational approaches and similar stigmergic concepts we derive a novel swarm intelligent algorithm. What strikes from these observations is that both eusocial insects as ant colonies and bacteria have similar natural mechanisms based on stigmergy in order to emerge coherent and sophisticated patterns of global collective behaviour. Keeping in mind the above characteristics we will present a simple model to tackle the collective adaptation of a social swarm based on real ant colony behaviors (*SSA algorithm*) for tracking extrema in dynamic environments and highly multimodal complex functions described in the well-know *DeJong* test suite. Then, for the purpose of comparison, a recent model of artificial bacterial foraging (*BFOA algorithm*) based on similar stigmergic features is described and analyzed. Final results indicate that the *SSA* collective intelligence is able to cope and quickly adapt to unforeseen situations even when over the same cooperative foraging period, the community is requested to deal with two different and contradictory purposes, while outperforming *BFOA* in adaptive speed. Results indicate that the present approach deals well in severe Dynamic Optimization problems.**

*Index Terms*—**Swarm Intelligence and Perception, Social Cognitive Maps, Social Foraging, Self-Organization, Distributed Search and Optimization in Dynamic Environments.**

## I. Introduction

SWARM Intelligence (SI) is the property of a system whereby the collective behaviors of (unsophisticated) entities interacting locally with their environment cause coherent functional global patterns to emerge. SI provides a basis with which it is possible to explore collective (or distributed) problem solving without centralized control or the provision of a global model (Stan Franklin, *Coordination without Communication*, talk at Memphis Univ., USA, 1996). The well-know bio-inspired computational paradigms know as ACO (*Ant Colony Optimization* algorithm [5]) based on trail formation via pheromone deposition / evaporation, and PSO (*Particle Swarm Optimization* [14]) are just two among many successful examples. Yet, and in what specifically relates to the biomimicry of these and other computational models, much more can be of useful employ, namely the social foraging behavior theories of many species, which can provide us with consistent hints to algorithmic approaches for the construction of social cognitive maps, self-organization [1,6], coherent swarm perception and intelligent distributed search, with direct applications in a high variety of social sciences and engineering fields [25→30]. In the present work, we will address the collective adaptation of a social community to a cultural (environmental, contextual) or informational dynamical landscape, represented here – for the purpose of different experiments – by several 3D mathematical functions that change over time. Our precise and final goal will be to keep track of extrema on those environments. For instance, typical applications of evolutionary optimization in static environments involve the approximation of the extrema of functions. On the contrary, for dynamic environments, the interest is not to locate the extrema but to follow it as closely as possible [12].

Flocks of migrating birds and schools of fish are familiar examples of spatial self-organized patterns formed by living organisms through social foraging. Such aggregation patterns are observed not only in colonies of organisms as simple as single-cell bacteria, as interesting as social insects like ants and termites as well as in colonies of multi-cellular vertebrates as complex as birds and fish but also in human societies [8]. Wasps, bees, ants and termites all make effective use of their environment and resources by displaying collective "swarm" intelligence. For example, termite colonies build nests with a complexity far beyond the comprehension of the individual termite, while ant colonies dynamically allocate labor to various vital tasks such as foraging or defense without any central decision-making ability [5]. Slime mould is another perfect example. These are very simple cellular organisms with limited motile and sensory capabilities, but in times of food shortage they aggregate to form a mobile slug capable of transporting the assembled individuals to a new feeding area. Should food shortage persist, they then form into a fruiting body that disperses their spores using the wind, thus ensuring the survival of the colony [18].

Vitorino Ramos, Carlos Fernandes and Agostinho Rosa, are with LaSEEB-ISR *Evolutionary Systems and BioMedical Eng. Lab.*, IST - Technical University of Lisbon (IST), Av. Rovisco Pais, 1, TN 6.21, 1049-001, Lisbon, PORTUGAL (corresponding author e-mails : vramos@laseeb.org, cfernandes@laseeb.org, acrosa@laseeb.org). Second author work was supported in part by FCT-PRAXIS XXI, *Ministério da Ciência, Tecnologia e Ensino Superior*, under a PhD fellowship. **Ajith Abraham** is with the School of Computer Science and Engineering, Chung-Ang University, Seoul, South Korea. (e-mail: *ajith.abraham@ieee.org* ).

New research suggests that microbial life can be even richer: highly social, intricately networked, and teeming with interactions. Bassler [2] and other researchers have determined that bacteria communicate using molecules comparable to pheromones, as ant colonies so often do. By tapping into this cell-to-cell network, microbes are able to collectively track changes in their environment, conspire with their own species, build mutually beneficial alliances with other types of bacteria, gain advantages over competitors, and communicate with their hosts - the sort of collective strategizing typically ascribed to bees, ants, and people, not to bacteria. Eshel Ben-Jacob [4] indicate that bacteria have developed intricate communication capabilities (e.g. quorum-sensing, chemotactic signalling and plasmid exchange) to cooperatively self-organize into highly structured colonies with elevated environmental adaptability, proposing that they maintain linguistic communication. Meaning-based communication permits colonial identity, intentional behaviour (e.g. pheromone-based courtship for mating), purposeful alteration of colony structure (e.g. formation of fruiting bodies), decision-making (e.g. to sporulate) and the recognition and identification of other colonies – features we might begin to associate with a bacterial social intelligence. Such a social intelligence, should it exist, would require going beyond communication to encompass unknown additional intracellular processes to generate inheritable colonial memory and commonly shared genomic context. Moreover, Eshel [3] argues that colonies of bacteria are able to communicate and even alter their genetic makeup in response to environmental challenges, asserting that the lowly bacteria colony is capable of computing better than the best computers of our time, and attributes to them properties of creativity, intelligence, and even self-awareness. These self-organizing distributed capabilities were also found in plants. Peak and co-workers [23] point out that plants may regulate their uptake and loss of gases by distributed computation – using information processing that involves communication between many interacting units (their *stomata*). As described, leaves have openings called stomata that open wide to let $CO_2$ in, but close up to prevent precious water vapour from escaping. Plants attempt to regulate their stomata to take in as much $CO_2$ as possible while losing the least amount of water. But they are limited in how well they can do this: leaves are often divided into patches where the stomata are either open or closed, which reduces the efficiency of $CO_2$ uptake. By studying the distributions of these patches of open and closed stomata in leaves of the cocklebur plant, Peak et al. [23] found specific patterns reminiscent of distributed computing. Patches of open or closed stomata sometimes move around a leaf at constant speed, for example. What's striking is that it is the same form of mechanism that is widely thought to regulate how ants forage. The signals that each ant sends out to other ants, by laying down chemical trails of pheromone, enable the ant community as a whole to find the most abundant food sources. Wilson [32] showed that ants emit specific pheromones and identified the chemicals, the glands that emitted them and even the fixed action responses to each of the various pheromones. He found that pheromones comprise a medium for communication among the ants, allowing fixed action collaboration, the result of which is a group behaviour that is adaptive where the individual's behaviours are not.

## II. SELF-ORGANIZATION AND STIGMERGY

Many structures built by social insects are the outcome of a process of self-organization [27,28], in which the repeated actions of the insects in the colony interact over time with the changing physical environment to produce a characteristic end state [11]. A major mediating factor is stigmergy [31], the elicitation of specific environment-changing behaviors by the sensory effects of local environment changes produced by previous and past behavior of the whole community. Stigmergy is a class of mechanisms that mediate animal-animal interactions through artifacts or via indirect communication, providing a kind of environmental synergy, information gathered from work in progress, a distributed incremental learning and memory among the society. In fact, the work surface is not only where the constituent units meet each other and interact, as it is precisely where a dynamical cognitive map could be formed, allowing for the embodiment of adaptive memory, cooperative learning and perception [25→30]. Constituent units not only learn from the environment as they can change it over time. Its introduction in 1959 by Pierre-Paul Grassé[1] made it possible to explain what had been until then considered paradoxical observations: In an insect society individuals work as if they were alone while their collective activities appear to be coordinated. The stimulation of the workers by the very performances they have achieved is a significant one inducing accurate and adaptable response. The phrasing of his introduction of the term is worth noting (translated to English in [11]):

*The coordination of tasks and the regulation of constructions do not depend directly on the workers, but on the constructions themselves.* **The worker does not direct his work, but is guided by it**. *It is to this special form of stimulation that we give the name Stigmergy (**stigma** - wound from a pointed object, and **ergon** - work, product of labor = stimulating product of labor).*

Keeping in mind the above characteristics (section I and II) we will present a simple model to tackle the collective adaptation of a social swarm based on real ant colony behaviors (*Swarm Search Algorithm* SSA - section III / results on section IV). Then, and for the purpose of comparison, a recent model of artificial bacterial foraging [22,17] (*Bacterial Foraging Optimization Algorithm* - BFOA) based on similar stigmergic features is described and analyzed (section V). Final results indicate that the SSA collective intelligence is able to cope and quickly adapt to unforeseen situations even when over the same cooperative foraging period, the community is requested to deal with two different and contradictory purposes, outperforming BFOA.

---

[1] Grassé, P.P.: La reconstruction du nid et les coordinations inter-individuelles chez *Bellicositermes natalensis* et *Cubitermes sp*. La théorie de la stigmergie : Essai d'interprétation des termites constructeurs. *Insect Sociaux* (1959), 6, 41-83.

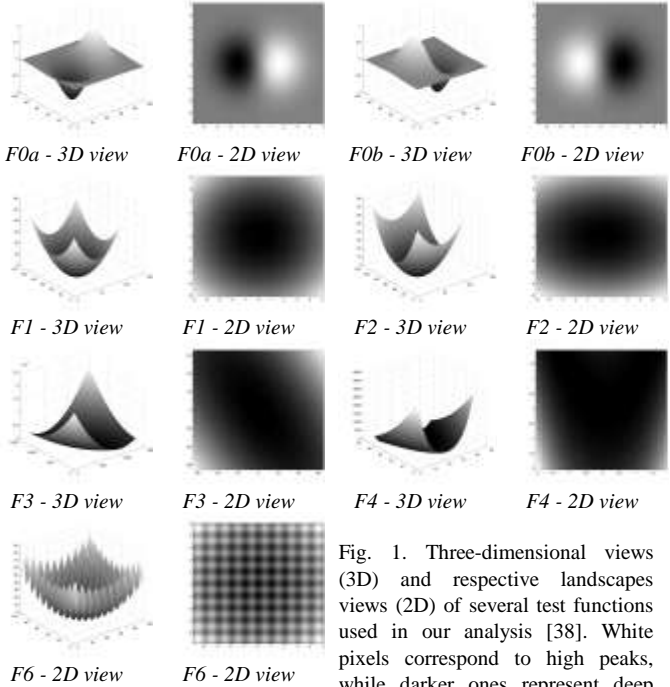

Fig. 1. Three-dimensional views (3D) and respective landscapes views (2D) of several test functions used in our analysis [38]. White pixels correspond to high peaks, while darker ones represent deep valleys (*F0-F4*) or holes (*F6*). Check table II in section 4.

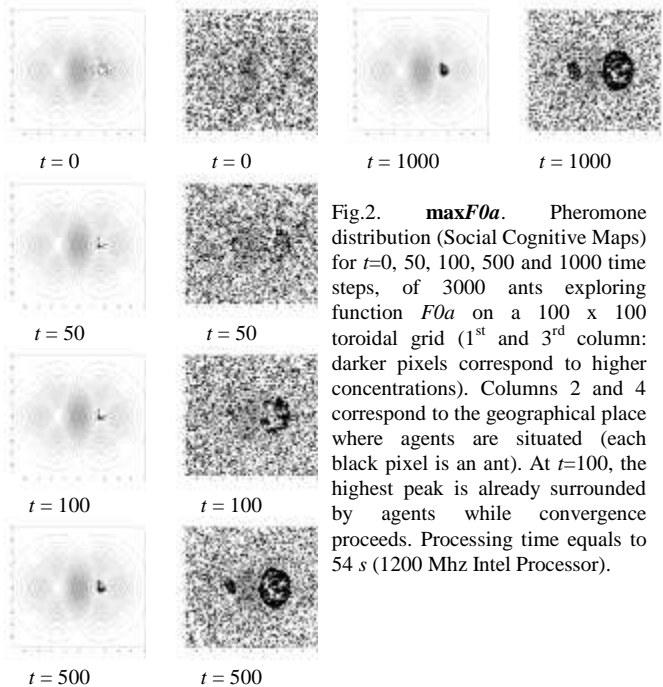

Fig.2. **max*F0a***. Pheromone distribution (Social Cognitive Maps) for *t*=0, 50, 100, 500 and 1000 time steps, of 3000 ants exploring function *F0a* on a 100 x 100 toroidal grid (1st and 3rd column: darker pixels correspond to higher concentrations). Columns 2 and 4 correspond to the geographical place where agents are situated (each black pixel is an ant). At *t*=100, the highest peak is already surrounded by agents while convergence proceeds. Processing time equals to 54 s (1200 Mhz Intel Processor).

### III. A SWARM MODEL FOR FORAGING IN DYNAMIC ENVIRONMENTS

As mentioned above, the distribution of the pheromone represents the memory of the recent history of the swarm (his social cognitive map), and in a sense it contains information which the individual ants are unable to hold or transmit [29]. There is no direct communication between the organisms but a type of indirect communication through the *pheromonal* field.

TABLE I
HIGH-LEVEL DESCRIPTION OF THE SWARM SEARCH ALGORITHM PROPOSED

```
/* Initialization */
For all agents do
   Place agent at randomly selected site
End For
/* Main loop */
For t = 1 to t_max do
   For all agents do
      /* According to Eqs. 1 and 2 (section 3) */
Compute W(σ) and P_ik
Move to a selected neighboring site not
occupied by other agent
   /* According to Eq. 3 (section 3) */
   Increase pheromone at site r:
               P_r= P_r+[η+p(Δ[r]/Δmax)]
   End For
   Evaporate pheromone by K, at all grid sites
End For
Print location of agents
Print pheromone distribution at all sites
/* Values of parameters used in experiments */
k = 0.015, η = 0.07, β=3.5, γ=0.2,
p = 1.9, t_max = 500, 600, 1000 or 1150 steps.
/* Useful references */
Check [25], [27], [7], [21] and [20].
```

In fact, ants are not allowed to have any local memory and the individual's spatial knowledge is restricted to local information about the whole colony pheromone density. In order to design this behaviour, one simple model was adopted [7], and extended due to specific constraints of the present proposal, in order to deal with 3D dynamic environments. As described by *Chialvo* and *Millonas*, the state of an individual ant can be expressed by its position *r*, and orientation $\theta$. Since the response at a given time is assumed to be independent of the previous history of the individual, it is sufficient to specify a transition probability from one place and orientation $(r,\theta)$ to the next $(r^*,\theta^*)$ an instant later. In previous works by Millonas [21,20], transition rules were derived and generalized from noisy response functions, which in turn were found to reproduce a number of experimental results with real ants. The response function can effectively be translated into a two-parameter transition rule between the cells by use of a pheromone weighting function (Eq.1):

$$W(\sigma)=\left(1+\frac{\sigma}{1+\gamma\sigma}\right)^{\beta} \qquad (1)$$

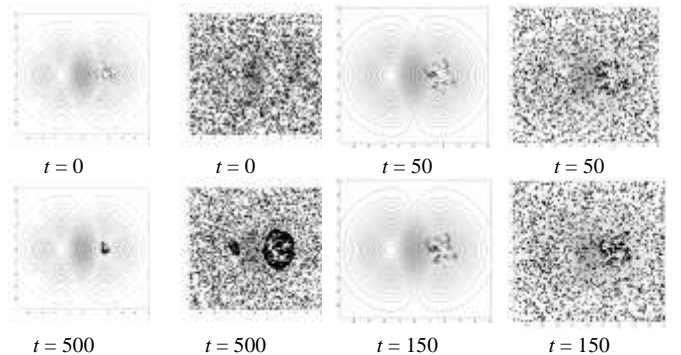

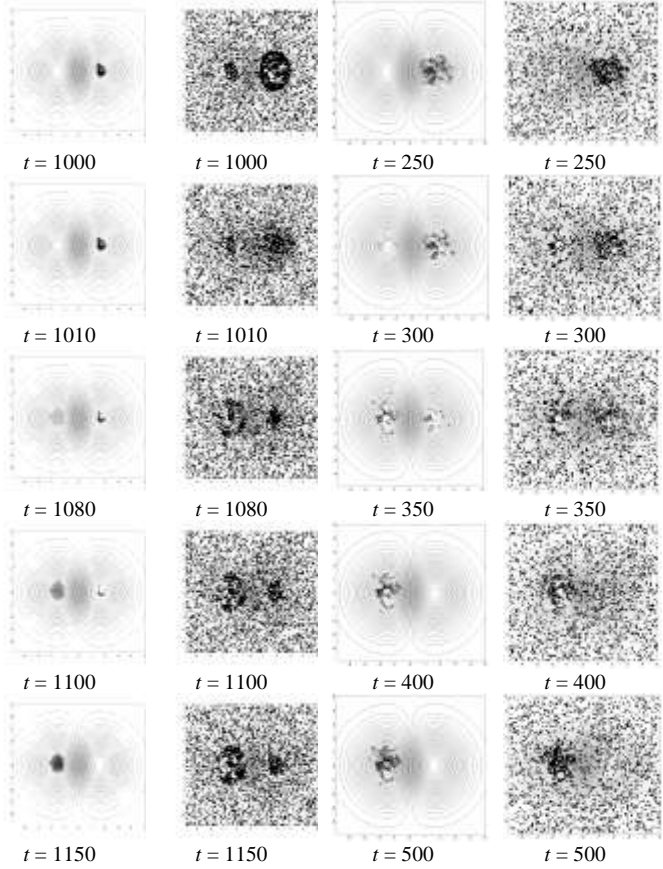

| | | | |
|---|---|---|---|
| $t = 1000$ | $t = 1000$ | $t = 250$ | $t = 250$ |
| $t = 1010$ | $t = 1010$ | $t = 300$ | $t = 300$ |
| $t = 1080$ | $t = 1080$ | $t = 350$ | $t = 350$ |
| $t = 1100$ | $t = 1100$ | $t = 400$ | $t = 400$ |
| $t = 1150$ | $t = 1150$ | $t = 500$ | $t = 500$ |

Fig. 3. **max$F0a$ => max$F0b$**. Social evolution from maximizing function $F0a$ to maximizing function $F0b$. In the first 1000 time steps the ant colony explores function $F0a$, while suddenly at $t=1001$, function $F0b$ is used as the new *habitat*. Pheromone distribution (Social Cognitive Maps) for $t = 0, 500, 1000, 1010, 1050, 1080, 1100$ and $1150$ time steps, of 3000 ants exploring function $F0a$ and $F0b$ on a 100 x 100 toroidal grid are shown. Already at $t=1010$, the old highest peak on the right suffers a radical erosion, on the presence of ants (they start to explore new regions).

Fig. 4. **max$F0a$ => min$F0a$**. Maximizing function $F0a$ during 250 time steps and then minimizing it for $t \geq 251$. Pheromone distribution (Social Cognitive Maps) for $t = 50, 150, 250, 300, 350, 400, 450$ and $500$ time steps, of 2000 ants exploring function $F0a$ on a 100 x 100 toroidal grid are shown. Already at $t=300$, the highest peak on the right suffers a radical erosion, on the presence of ants starting to explore new regions. As time passes the majority of the colony moves to the deep valley, on the left. Parameters are different from those used in Figs. 2-3 (check table III).

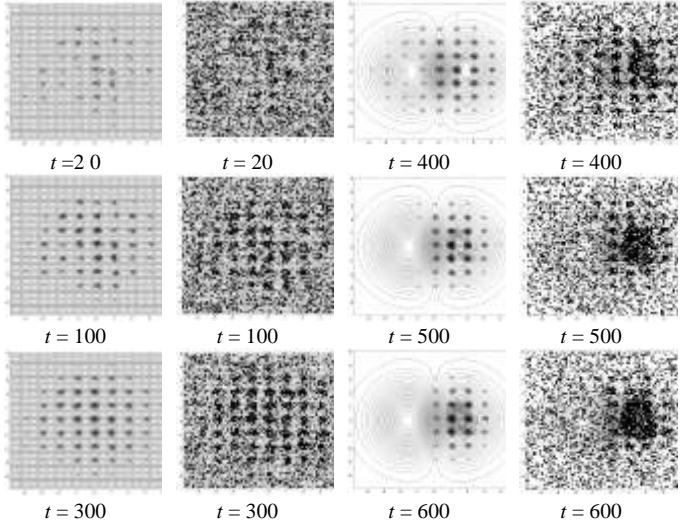

| | |
|---|---|
| $t = 20$ | $t = 20$ |
| $t = 100$ | $t = 100$ |
| $t = 300$ | $t = 300$ |

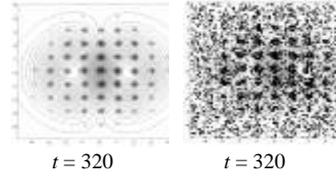

| | |
|---|---|
| $t = 320$ | $t = 320$ |
| $t = 400$ | $t = 400$ |
| $t = 500$ | $t = 500$ |
| $t = 600$ | $t = 600$ |

Fig. 5. **min$F6$ => max$F0a$**. Minimizing function $F6$ during 300 time steps and then maximizing function $F0a$ for $t \geq 301$. Pheromone distribution (Social Cognitive Maps) for $t = 20, 100, 300, 320, 400, 500,$ and $600$ time steps, of 3000 ants exploring function $F6$ and $F0a$ on a 100 x 100 toroidal grid are shown. Parameters are different from those used in Figs. 2-3 (check table III).

This equation measures the relative probabilities of moving to a cite $r$ (in our context, to a cell in the grid *habitat*) with pheromone density $\sigma(r)$. The parameter $\beta$ is associated with the osmotropotaxic sensitivity, recognised by Wilson [32] as one of two fundamental different types of ant's sense-data processing. *Osmotropotaxis*, is related to a kind of instantaneous pheromonal gradient following, while the other, *klinotaxis*, to a sequential method (though only the former will be considered in the present work as in [7]). Also it can be seen as a physiological inverse-noise parameter or gain. In practical terms, this parameter controls the degree of randomness with which each ant follows the gradient of pheromone. On the other hand, $1/\gamma$ is the sensory capacity, which describes the fact that each ant's ability to sense pheromone decreases somewhat at high concentrations.

$$P_{ik} = \frac{W(\sigma_i)w(\Delta_i)}{\sum_{j/k} W(\sigma_j)w(\Delta_j)} \qquad (2)$$

$$T = \eta + p\frac{\Delta[i]}{\Delta_{max}} \qquad (3)$$

In addition to the former equation, there is a weighting factor $w(\Delta\theta)$, where $\Delta\theta$ is the change in direction at each time step, i.e. measures the magnitude of the difference in orientation. As an additional condition, each individual leaves a constant amount $\eta$ of pheromone at the cell in which it is located at every time step $t$. This pheromone decays at each time step at a rate $k$. Then, the normalised transition probabilities on the lattice to go from cell $k$ to cell $i$ are given by $P_{ik}$ (Eq. 2, [7]), where the notation $j/k$ indicates the sum over all the surrounding cells $j$ which are in the local neighbourhood of $k$. $\Delta_i$ measures the magnitude of the difference in orientation for the previous direction at time $t$-1. That is, since we use a neighbourhood composed of the cell and its eight neighbours, $\Delta_i$ can take the discrete values 0 through 4, and it is sufficient to assign a value $w_i$ for each of these changes of direction. Chialvo et al. used the weights of $w_0 =1$ (same direction), $w_1 =1/2$, $w_2 =1/4$, $w_3 =1/12$ and $w_4 =1/20$ (U-turn). In addition, coherent results were found for $\eta=0.07$ (pheromone deposition rate), $k=0.015$ (pheromone evaporation rate), $\beta=3.5$ (osmotropotaxic sensitivity) and $\gamma=0.2$ (inverse of sensorycapacity), where the emergence of well defined networks of trails were possible. Except when indicated, these values will remain in the following framework. As an additional condition, each individual leaves a constant amount

$\eta$ of pheromone at the cell in which it is located at every time step $t$. Simultaneously, the pheromone evaporates at rate $k$, i.e., the pheromonal field will contain information about past movements of the organisms, but not arbitrarily in the past, since the field *forgets* its distant history due to evaporation in a time $\tau \cong 1/k$. As in past works, toroidal boundary conditions are imposed on the lattice to remove, as far as possible any boundary effects (e.g. one ant going out of the grid at the south-west corner, will probably come in at the north-east corner).

In order to achieve emergent and *autocatalytic* mass behaviours around specific extrema locations (e.g., peaks or valleys) on the *habitat*, instead of a constant pheromone deposition rate $\eta$ used in [7], a term not constant is included. This upgrade can significantly change the expected ant colony cognitive map (pheromonal field). The strategy follows an idea implemented earlier by Ramos [25,26], while extending the Chialvo model into digital image habitats, aiming to achieve a collective perception of those images by the end product of swarm interactions. The main differences to the Chialvo work is that ants, now move on a 3D discrete grid, representing the functions which we aim to study (fig. 1) instead of a 2D *habitat*, and the pheromone update takes in account not only the local pheromone distribution as well as some characteristics of the cells around one ant. In here, this additional term should naturally be related with specific characteristics of cells around one ant, like their altitude ($z$ value or function value at coordinates $x,y$), having in mind our present aim. So, our pheromone deposition rate $T$, for a specific ant, at one specific cell $i$ (at time $t$), should change to a dynamic value ($p$ is a constant = 1.93) expressed by equation 3. In this equation, $\Delta_{max} = | z_{max} - z_{min} |$, being $z_{max}$ the maximum altitude found by the colony so far on the function *habitat*, and $z_{min}$ the lowest altitude. The other term $\Delta[i]$ is equivalent to (if our aim is to minimize any given landscape): $\Delta[i] = | z_i - z_{max} |$, being $z_i$ the current altitude of one ant at cell $i$. If on the contrary, our aim is to maximize any given landscape, then we should instead use $\Delta[i] = | z_i - z_{min} |$. Finally, please notice that if our landscape is completely flat, results expected by this extended model will be equal to those found by Chialvo and Millonas in [7], since $\Delta[i]/\Delta_{max}$ equals to zero. In this case, this is equivalent to say that only the swarm pheromonal field is affecting each ant choices, and not the *environment* - i.e. the expected network of trails depends largely on the initial random position of the colony, and in trail clusters formed in the initial configurations of pheromone. On the other hand, if this environmental term is added a stable and emergent configuration will appear which is largely independent on the initial conditions of the colony and becomes more and more dependent on the nature of the current studied *landscape* itself. As specified earlier, the environment plays an active role, in conjunction with continuous positive and negative feedbacks provided by the colony and their pheromone, in order to achieve a stable emergent pattern, memory and distributed learning by the community [29].

TABLE II
CLASSICAL TEST FUNCTIONS USED IN OUR ANALYSIS FROM *MATLAB* [24]

| Function ID | Equation |
|---|---|
| F0a | $f_{0a}(x) = x_1 . e^{-0.2 \sum_{i=1}^{n} x_i^2}$ |
| F0b | $f_{0b}(x) = -x_1 . e^{-0.2 \sum_{i=1}^{n} x_i^2}$ |
| F1 | $f_1(x) = \sum_{i=1}^{n} x_i^2$ |
| F2 | $f_{1a}(x) = \sum_{i=1}^{n} i.x_i^2$ |
| F3 | $f_{1b}(x) = \sum_{i=1}^{n} \left( \sum_{j=1}^{i} x_j \right)^2$ |
| F4 | $f_2(x) = \sum_{i=1}^{n-1} 100.(x_{i+1} - x_i^2)^2 + (1 - x_i)^2$ |
| F5 | $f_6(x) = 10.n + \sum_{i=1}^{n} (x_i^2 - 10.\cos(2.\pi.x_i))$ |
| F6 | $f_7(x) = \sum_{i=1}^{n} -x_i . \sin(\sqrt{|x_i|})$ |

## IV. EXPERIMENTAL SETUP AND RESULTS

In order to test the dynamical behaviour of this new *Swarm Search* algorithm presented earlier in section 3 (pseudo-code in table I), we have used classical test functions (table II) drawn from the literature in Genetic Algorithms, Evolutionary strategies and global optimization [24], several of them graphically accessible in fig. 1. Function *F0a* represents one deep valley and one peak, while *F0b* his the opposite. Function *F1* represents *De Jong*'s function 1 and his one of the simplest. It is continuous, convex and unimodal; $x_i$ is in the interval [- 5.12; 5.12] and the global minimum is at $x_i$=0. Function *F2* represents an axis parallel hyper-ellipsoid similar to *De Jong*'s function 1. It is also know as the weighted sphere model. Again it is continuous, convex and unimodal in the interval $x_i \rightarrow$ [-5.12; 5.12], with global minimum at $x_i$=0. Function *F3* represents an extension of the axis parallel hyper-ellipsoid (*F2*), also know as *Schwefel*'s function 1.2. With respect to the coordinate axes this function produces rotated hyper-ellipsoids; $x_i$ is in the interval [-65.536; 65.536] and the global minimum is at $x_i$=0. Likewise *F2*, it is continuous, convex and unimodal. Function *F4* represents the well-know *Rosenbrock*'s valley or *De Jong*'s function 2. *Rosenbrock*'s valley is a classic optimization problem. The global optimum is inside a long, narrow, parabolic shaped flat valley. To find the valley is trivial, however convergence to the global optimum is difficult and hence this problem has been repeatedly used in assess the performance of optimization algorithms; $x_i$ is in the interval [-2.048; 2.048] and the global minimum is at $x_i$=0. Function *F5* represents the *Rastrigin*'s function 6. This function is based on *De Jong*'s function 1 with the addition of cosine modulation to produce many local minima. Thus, the test function is highly multimodal. However, the location of the minima are regularly distributed. As in *F1*, $x_i$ is in the interval [-5.12; 5.12] and the global minimum is at $x_i$=0. Finally, *F6* represents *Schwefel*'s function 7, being deceptive in that the global minimum is geometrically distant, over the parameter space, from the next best local minima. Therefore, the search algorithms are potentially prone to convergence in the wrong direction; $x_i$ is in the interval [-500; 500] and the global minimum is at $x_i$=420,9687 while $f(x)=n.418,9829$. In our tests, $n$=2. Within this specific framework we have produced several run tests using different test functions, some of which are presented here trough figures 2 to 5. The parameters used are shown on table 3. The simplest test was the first one (fig.2) where we forced the colony to search for the maximal peak in function *F0a*, during 1000 time steps. The other tests were harder, that is dynamic, since they include not only different purposes simultaneously (maximizing and minimizing), tracking different extrema, as well as different landscapes that changed dynamically on intermediate swarm search stages (e.g., fig. 3, 4 and 5).

TABLE III
PARAMETERS USED FOR DIFFERENT TEST SETS

| Fig. | N ants | $t_{max}$ | k | $\eta$ | $\beta$ | $\gamma$ | p |
|---|---|---|---|---|---|---|---|
| 2 | 3000 | 1000 | 0.015 | 0.07 | 3.5 | 0.2 | 1.93 |
| 3 | 3000 | 1150 | 0.015 | 0.07 | 3.5 | 0.2 | 1.93 |
| 4 | 2000 | 500 | 1.000 | 0.10 | 3.5 | 0.2 | 1.90 |
| 5 | 3000 | 600 | 1.000 | 0.01 | 3.5 | 0.2 | 1.90 |

## V. SWARM SEARCH VERSUS BACTERIAL FORAGING ALGORITHMS

In order to further analyze the collective behavior of the present proposal, we performed a comparison between the ant-like *Swarm Search Algorithm* (SSA) and the *Bacterial Foraging Optimization Algorithm* (BFOA), on the dominion of function optimization. BFOA was selected since it represents an earlier proposal for function optimization as well based on natural foraging capacities. Presented by Passino at *IEEE Control Systems Magazine* in 2002 [22] and later that year in the *Journal of Optimization Theory and Applications* [17], the author for the purpose of a simple but powerful illustrative example, used his algorithm to find the minimum of two complex functions $J_{cc}$, described in [22], page 60. Further material, as the MATLAB code of his algorithm and the tri-dimensional functions experimented, can also be found on the web address of a recent book from the same author (*Biomimicry for Optimization, Control and Automation*, Springer-Verlag, London, UK, 2005), at *http://www.ece.osu.edu/ ~passino/ICbook/ ic_index.html*. Passino uses $S$=50 bacteria-based agents, during four generations. In each generation, and has a requirement of his algorithm, each agent enters a chemotaxis loop (see page 61 [22]), performing $N_c$=100 chemotactic (foraging) steps.

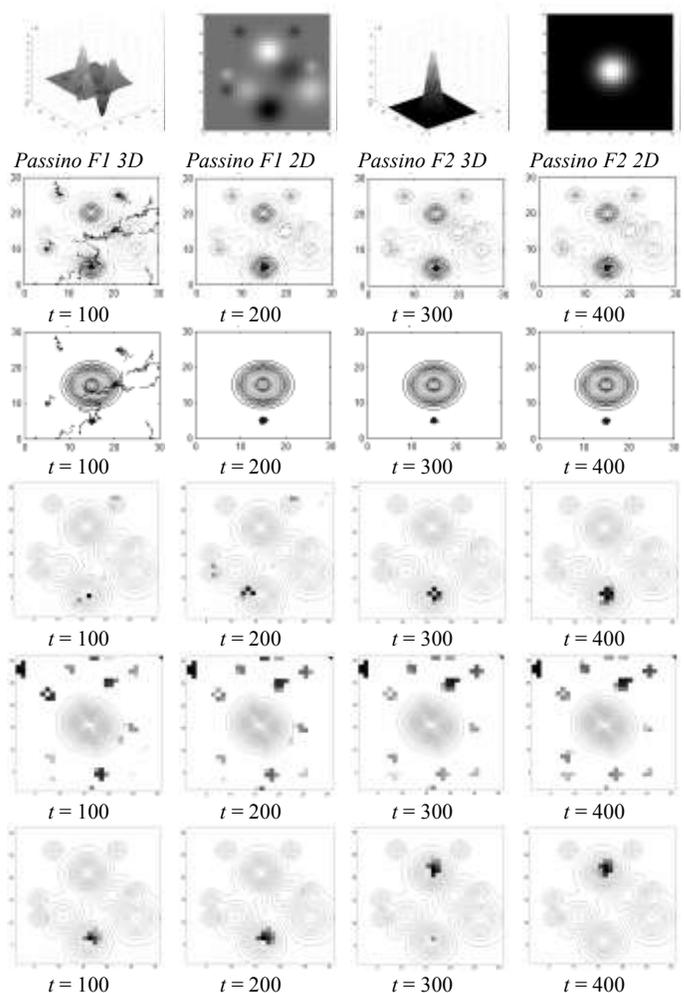

Fig. 6. In the first row the test functions used by Passino [22,17]. In the second and third rows, BFOA minimizing results respectively for *F1* and *F2*. The graphics show the bacterial motion trajectories (using 50 bacteria-like agents). In the fourth and fifth rows, SWARM-SEARCH algorithm (SSA) minimizing results respectively for *F1* and *F2*, and for the same foraging time period. The graphics shows the pheromone distribution. In the last row, SSA is requested to deal with two contradictory goals, i.e. to minimize *F1* and then to maximize it. In all these tests, SSA has used 50 ant-like agents. Check main text for the parameters used. Habitat size equals 2 x [0,30].

Thus the algorithm – for the precise application – runs for $t$=400 time steps, which make us believe that a fair comparison can be make in regard of the parameter values we use. The two functions represent what Passino designates by *nutrient concentration landscapes* (see fig. 6, first row – the web address also contains his MATLAB code used in the two functions, where *Nutrientsfunc.m* and *Nutrientsfunc1.m* are represented by different weights). His function *F2* (*Nutrientsfunc1.m*) has a zero value at [15,15] and decreases to successively more negative values as you move away from that point, reaching a *plateau* with the same value. Moreover, and for the purpose of discrete function optimization, Passino [22,17] represented both functions by a discrete lattice (as well as us in our past tests) with a size of 30 x 30 cells over the optimization domain (each cell has a correspondent $z$ or $J_{cc}$ value). For these reasons and in order to keep a coherent comparison, we shall use 50 ant-like agents in our SSA, on a 30 x 30 tri-dimensional *habitat*, for $t$=400 time steps, on both functions. We then run 3 tests. The first is requested to minimize Passino's function *F1*. The second test is requested to minimize Passino's function *F2*. Finally, and in order to prove the highly adaptive features of our model, we requested SSA to deal with two contradictory goals, i.e. to minimize *F1* and then to maximize it, over the same period of 400 time steps. As visible, SSA quickly adapts to the different purposes. Over function *F1*, the pheromone concentration is already intensely allocated at the right point at $t=100$ (and not in other areas), while BFOA, at this moment, still explores different regions on the optimization domain. Over function *F2*, the swarm quickly separates in different foraging groups, since there are a large number of points with the same minimal value. Finally over function *F1* again, in the final test (last row – fig. 6), SSA is able to process two different demands (minimization followed by maximization) over the same foraging time period that BFOA uses for *F1* minimization. The parameters used in our experiments follows: $N_{ants}=50$, $t_{max}=400$, $k=1$ (pheromone evaporation rate), $\eta=0.1$ (pheromone deposition rate), $\beta=7$ (this parameter controls how ants follow the pheromone gradient), $\gamma=0.2$, and $p=1.9$. Exception made for test 1, where $\beta=6$.

## VI. CONCLUSIONS

Evolution of mass behaviours on time are difficult to predict, since the global behaviour is the result of many part relations operating in their own local neighbourhood. The emergence of network trails in ant colonies, for instance, are the *product* of several simple and local interactions that can evolve to complex patterns, which in some sense translate a meta-behaviour of that swarm [29]. Moreover, the translation of one kind of low-level (present in a large number) to one meta-level is minimal. Although that behaviour is specified (and somehow constrained), there is minimal specification of the mechanism required to generate that behaviour; global behaviour evolves from the many relations of multiple simple behaviours, without global coordination (i.e. from local interactions to global complexity. There is some evidence that our brain as well as many other complex systems, operates in the same way, and as a consequence collective perception capabilities could be derived from emergent properties, which cannot be neglected in any pattern search algorithm. These systems show in general, interesting and desirable features as *flexibility* (e.g. the brain is able to cope with incorrect, ambiguous or distorted information, or even to deal with unforeseen or new situations without showing abrupt performance breakdown) or *versability*, *robustness* (keep functioning even when some parts are locally damaged), and they operate in a massively parallel fashion. Present results point to that type of interesting features. Although the current model is far from being consistent with real ones, since only some type of real mechanisms were considered, swarm pheromonal fields reflect some convergence towards *the identification* of a common goal in a purely decentralized form. Moreover, the present model shows important adaptive capabilities, as in the presence of sudden changes in the *habitat* - our test landscapes (fig. 1). Even if the model is able to quickly adapt to one specific environment, evolving from one empty pheromonal field, *habitat* transitions point that, the whole system is able to have some memory from past environments (i.e. convergence is more difficult after *learning* and *perceiving* one past *habitat*). On the other hand this feature can have some advantage, for instance in the case where the original or similar environments are back in place. This emerged feature of *résistance*, is somewhat present in many of the natural phenomena that we find today in our society. In a certain sense, the distribution of pheromone represents the collective solutions found so far (memory, risk avoidance, exploitation behavior), while evaporation enables the system to adapt (tricks a decision, explorative behavior), not only as in normal situations (a complex but static search environment), as well as when the landscape suddenly changes, moving the colony's new target to a new unexplored region and keep tracking of it. One crucial aspect observed here, as noted in the past by Langton [16] and present in many complex systems, only at the right intermediary regime, in here between contradictory behaviors of exploration and exploitation, the swarm is able to quickly converge.

The recognizable results indicate that the collective intelligence is able to cope and quickly adapt to unforeseen situations even when over the same cooperative foraging period, the community is requested to deal with two different and contradictory purposes. All these above mentioned aspects show how vital can be the study of social foraging for the development of new distributed search algorithms, and the construction of social cognitive maps, with interesting properties in collective memory, collective decision-making and swarm-based pattern detection and recognition.

But the work could have important consequences in other areas. Perhaps, one of the most valuable relations to explore is that of social foraging and evolution. For two reasons; First, as described by Passino [22], natural selection tends to eliminate animals with poor "foraging strategies" (methods for locating, handling, and ingesting food) and favor the propagation of genes of those animals that have successful foraging strategies since they are more likely to enjoy reproductive success (they obtain enough food to enable them to reproduce). Logically, such evolutionary principles have led scientists in the field of foraging theory to hypothesize that it is appropriate to model the activity of foraging as an optimization process: A foraging animal takes actions to maximize the energy obtained per unit time spent foraging, in the face of constraints presented by its own physiology and by the environment.

Second, because there is an increasing recognition that natural selection and self-organization work hand in hand to form evolution, as defended by Kauffmann [13]. For example, anthropologist Jeffrey McKee [19,14] has described the evolution of human brain as a self-organizing process. He uses the term autocatalysis to describe how the design of an organism's features at one point in time affects or even determines the kinds of designs it can change into later. For example the angle of the skull on the top of the spine left some extra space for the brain to expand. Thus the evolution of the organism is determined not only by selection pressures but by constraints and opportunities offered by the structures that have evolved so far. Also, and back again in what regards the evolution of collectives, it is known that during the evolution of life, there have been several transitions in which individuals began to cooperate, forming higher levels of organization and sometimes losing their independent reproductive identity (insect societies are one example). Several factors that confer evolutionary advantages on higher levels of organization have been proposed, such as *Division of Labor* and *Increased Size*. But recently, a new third factor was added: *Information Sharing* [15]. Lachmann et al., illustrate with a simple model how information sharing can result in individuals that both receive more information about their environment and pay less for it. Being social foraging essentially a self-organized phenomenon, the study of computational foraging embedded with GA (Genetic Algorithm) like natural selection can much probably enhance our understanding on the detailed forms of the hypothetical equation: *Evolution = Natural Selection + Self-Organization*, and in the precise role of each "variable". As an example, current work in the same area [10], include the research of variable population size swarms, as used similarly in Evolutionary Computation [9], where each individual can have a probability of making a child, as well to die, depending on his *accumulated* versus *spent* energetic resources. The system as a whole, then proceeds on the search space as a kind of distributed evolutionary swarm. Finally and in parallel, an effort is being made in order to understand the societal memory and his speed on tracking extrema over dynamic environments using self-regulatory swarms based on the present model [30,10,29].

REFERENCES

[1] Bak, P., How Nature Works – The Science of Self-Organized Criticality, Springer-Verlag, 1996.
[2] Bassler, B.L, "Small Talk: Cell-to-Cell Communication in Bacteria", Cell, Vol. 109, pp. 421-424, May 2002.
[3] Ben-Jacob, E., Shochet, O., Tenenbaum, A., Cohen, I., Czirók, A., Vicsek, T., "Generic Modelling of Cooperative Growth in Bacterial Colonies", Nature, 368, pp. 46-49, 1994.
[4] Ben-Jacob, E., Becker, I., Shapira, Y., Levine, H., "Bacterial Linguistic Communication and Social Intelligence", Trends in Microbiology, Vol. 12/8, pp. 366-372, 2004.
[5] Bonabeau, E., Dorigo, M., Theraulaz, G., Swarm Intelligence: From Natural to Artificial Systems, Santa Fe Institute in the Sciences of Complexity, Oxford Univ. Press, New York, Oxford, 1999.
[6] Camazine, S., Deneubourg, J.-L., Franks, N.R., Sneyd, J., Theraulaz, G., Bonabeau, E., Self-Organization in Biological Systems, Princeton Studies in Complexity, Princeton University Press, 2001.
[7] Chialvo, D.R., Millonas, M.M., "How Swarms build Cognitive Maps", In Steels, L. (Ed.): The Biology and Technology of Intelligent Autonomous Agents, 144, NATO ASI Series, 439-450, 1995.
[8] Chowdhury, D., Nishinari, K., Schadschneider, A., "Self-Organized Patterns and Traffic Flow in Colonies of Organisms: from Bacteria and Social Insects to Vertebrates", special issue on Pattern Formation, in Phase Transitions, Taylor and Francis, vol. 77, 601, 2004.
[9] Fernandes, C., Rosa, A.C., "Study on Non-random Mating and Varying Population Size in Genetic Algorithms using a Royal Road Function", IEEE CEC´01, Proc. of the 2001 IEEE Congress on Evolutionary Computation, pp. 60-66, 2001.
[10] Fernandes, C., Ramos, V, Rosa, A.C., "Varying the Population Size of Artificial Foraging Swarms on Time Varying Landscapes", to appear in ICANN-05, Int. Conf. on Artificial Neural Networks, Springer-Verlag, LNCS Series, Warsaw, Poland, Sept. 11-15, 2005.
[11] Holland, O., Melhuish, C.: Stigmergy, "Self-Organization and Sorting in Collective Robotics", Artificial Life, Vol. 5, n. 2, MIT Press, 173, 1999.
[12] Huang, C.-F., Rocha, L.M., "Tracking Extrema in Dynamic Environments using a Coevolutionary Agent-based Model of Genotype Edition", in GECCO-05, Genetic and Evolutionary Computation Conf., Washington, D.C., USA, 25-29 June, 2005.
[13] Kauffmann, S.A., The Origins of Order: Self-Organization and Selection in Evolution, New York: Oxford University Press, 1993.
[14] Kennedy, J. Eberhart, Russel C. and Shi, Y., Swarm Intelligence, Academic Press, Morgan Kaufmann Publ., San Diego, London, 2001.
[15] Lachmann, M., Sella, G., Jablonka, E., "On Information Sharing and the Evolution of Collectives", Proc. of the Royal Society: Biological Sciences, 267, pp. 1265-1374, 2000.
[16] Langton, C.G., "Computation at the Edge of Chaos", Physica D, 42, pp. 12-37, 1990.
[17] Liu, Y., Passino, K.M., "Biomimicry of Social Foraging Bacteria for Distributed Optimization: Models, Principles, and Emergent Behaviors", Journal of Optimization Theory and Applications, Vol. 115, nº3, pp. 603-628, Dec. 2002.
[18] Maree, A.F.M., Hogeweg, P., "How Amoeboids Self-Organize into a Fruiting Body: Multicellullar Coordination in Dictyostelium discoideum", PNAS, vol. 98, nº 7, pp. 3879-3883, 2001.
[19] McKee, J.K., The Riddled Chain: Change, Coincidence, and Chaos in Human Evolution, Piscataway, NJ: Rutjers University Press, 2000.
[20] Millonas, M.M., "A Connectionist-type model of Self-Organized Foraging and Emergent Behavior in Ant Swarms", J. Theor. Biol., nº 159, 529, 1992.
[21] Millonas, M.M., "Swarms, Phase Transitions and Collective Intelligence", In Langton, C.G. (Ed.): Artificial Life III, Santa Fe Institute, Studies in the Sciences of Complexity, Vol. XVII, Addison-Wesley, Reading, Massachusetts, 417-445, 1994.
[22] Passino, K.M., "Biomimicry of Bacterial Foraging for Distributed Optimization and Control", IEEE Control Systems Magazine, pp. 52-67, June 2002.
[23] Peak, D.A., West, J.D., Messinger, S.M., Mott, K.A., "Evidence for Complex, Collective Dynamics and Emergent, Distributed Computation in Plants", PNAS, Proc. of the National Academy of Sciences, USA, 101, pp. 918-922, 2004.
[24] Pohlheim, H, "Genetic Algorithm MATLAB Toolbox Test Functions", MATLAB reference manual, version 1.2, Mathworks, 1997.
[25] Ramos, V., Almeida, F., "Artificial Ant Colonies in Digital Image Habitats: A Mass Behavior Effect Study on Pattern Recognition", In Dorigo, M., Middendorf, M., Stuzle, T. (Eds.): From Ant Colonies to Artificial Ants - 2nd Int. Wkshp on Ant Algorithms, 113-116, 2000.
[26] Ramos, V., "On the Implicit and on the Artificial - Morphogenesis and Emergent Aesthetics in Autonomous Collective Systems", in ARCHITOPIA Book, Art, Architecture and Science, Institut D'Art Contemporain, J.L. Maubant et al. (Eds.), pp. 25-57, Chapter 2, ISBN 2905985631 – EAN 9782905985637, France, Feb. 2002.
[27] Ramos,V., Merelo, Juan J., "Self-Organized Stigmergic Document Maps: Environment as a Mechanism for Context Learning", in AEB'2002 – 1st Spanish Conf. on Evolutionary and Bio-Inspired Algorithms, E. Alba, F. Herrera, J.J. Merelo et al. (Eds.), pp. 284-293, Centro Univ. de Mérida, Mérida, Spain, 6-8 Feb. 2002.
[28] Ramos, V., Abraham, A., "Evolving a Stigmergic Self-Organized Data-Mining", in ISDA-04, 4th Int. Conf. on Intelligent Systems, Design and Applications, Budapest, Hungary, ISBN 963-7154-30-2, pp. 725-730, August 26-28, 2004.
[29] Ramos, V., Fernandes, C., Rosa, A.C., "Social Cognitive Maps, Swarm Collective Perception and Distributed Search on Dynamic Landscapes", to appear in *Brains, Minds & Media* – Journal of New Media in Neural and Cognitive Science, NRW, Germany, 2005.
[30] Ramos, V., Fernandes, C., Rosa, A.C., "Societal Memory and his Speed on Tracking Extrema over Dynamic Environments using Self-Regulatory Swarms", invited paper at NiSIS-05, 1st European Symp. on Nature-inspired Smart Information Systems, Portugal, 3-5 Oct., 2005.
[31] Theraulaz, G., Bonabeau, E., "A Brief History of Stigmergy", Artificial Life, Vol. 5, n. 2, MIT Press, 97-116, 1999.
[32] Wilson, E.O., The Insect Societies, Cambridge, MA., Belknap Press, 1971.